\documentclass[namedreferences]{solarphysics}
\usepackage[hyperref,optionalrh,solaromanenum]{spr-sola-addons} 
\usepackage{graphicx}                    
\usepackage{color}                       
\usepackage{breakurl}                         

\begin{document}

\begin{article}

\begin{opening}

\title{{\v S}vestka's Research Then and Now\\ {\small Invited Review}}

%
\author{H.~S.~\surname{Hudson}$^{1,2}$\sep
        }

%
\runningauthor{H.~Hudson}
\runningtitle{Then and Now}

%
  \institute{$^{1}$ SSL, UC Berkeley, 7 Gauss Way, Berkeley CA 94720 USA\\
                  email: \url{hhudson@ssl.berkeley.edu}\\
           $^{2}$ University of Glasgow
             }

\begin{abstract}
Zden{\v e}k {\v S}vestka's research work influenced many fields of solar physics, especially in the area of flare research.
In this article I take five of the areas that particularly interested him and assess them in a ``then and now'' style. 
His insights in each case were quite sound, although of course in the modern era we have learned things that he could not readily have envisioned.
His own views about his research life have been published recently in this journal, to which he contributed so much, and his memoir contains much additional scientific and personal information \citep{2010SoPh..267..235S}.
\end{abstract}

%

\end{opening}

\section{Introduction}

In a quasi-historical, quasi-scientific review such as this, one must expect a fairly ignorant account that is heavily biased to the author's own knowledge base and working concepts.
Since the author's own research career only began some two decades after Zden{\v e}k's, and since  we had not actually met until the late 1960s,
the literature trail provides the main material for the background, along with the author's speculations about the scientific environment prior to that time.
Thus the reader should be wary of a certain degree of mismatch (knowledge and opinion) between Zden{\v e}k and myself, but in general we have certainly been interested in the same topics, and in particular the still-fascinating white-light flare phenomenon.

I have structured this as a ``then and now'' summary of specific topics to which Zden{\v e}k made important early contributions, and recommend that any interested reader also read Zden{\v e}k's own memoir \citep{2010SoPh..267..235S} for further insight especially into the ``then''  aspects. 
The five specific topics (in Sections 2--6: optical spectroscopy, proton flares, loop-prominence systems, the Flare Build-up Study, and ``giant arches'') overlap considerably and mainly provide a framework in time for the discussion.
No doubt one thing led to another for Zden{\v e}k's research interest, as it will, and the topics have considerable overlap because of the broadly interconnected dynamics of the components of the overall flare process.
He had a prescient interest in ``proton flares,'' which nowadays we would call Solar Particle Events (jargonized to SPEs, which consist of Solar Energetic Particles, or SEPs).
This interest predated the discovery of coronal mass ejections (CMEs), now known to play a vital role in particle acceleration, and we will pick up the historical trail below.

A solar flare is a complicated set of phenomena that reflect sudden energy release in the solar atmosphere \cite[\textit{e.g.}][]{1976sofl.book.....S,2011SSRv..159...19F}.
The effects range from the photosphere (``white-light flares'') to the distant heliosphere (``Interplanetary Coronal Mass Ejections,'' or ICMEs). 
Flares can occur without detectable CMEs, and sometimes \textit{vice versa}, with the ``stealth CMEs'' exemplifying the latter \citep[\textit{e.g.}][]{1998GeoRL..25.2469W,2009ApJ...701..283R}.
The latter typically come from filament channels remote from active regions.
The white-light emission of a solar flare reflects radiation emitted at the photosphere itself, or even from  slightly below it \citep{2004ApJ...607L.131X}.
Flare effects indeed extend into the deep interior, as known via seismic signatures \cite{1998Natur.393..317K}.
The energy that is to be released during one of these sudden transients comes from slowly-accumulated stresses in the magnetic field, according to general consensus, and indeed alternative possible sources for this energy seem implausible \citep[\textit{e.g.}][]{2007ASPC..368..365H}.

High-resolution spectroscopic observations of solar flares (Section~\ref{sec:wlf}), Zden{\v e}k's main interest in the early decades, has returned to the forefront now, both because technology has advanced considerably beyond photographic techniques and because the limitations imposed by Earth's atmosphere have disappeared. 
As revealed with crystal clarity by the original observations of solar flares \citep{1859MNRAs..20...13C,1859MNRAs..20...16H} because of the large radiated energy, the lower solar atmosphere plays a fundamental role in flare development.
Nowadays, not only do we have the broader coverage afforded by observatories in space, but are also seeing a remarkable development of high-resolution ground-based observatories with much-improved optical properties and data capabilities.
For this reason the open questions from Zden{\v e}k's early work may soon find some answers, and it is timely to consider how these questions arose.

As a personal note about this paper: I learned a great deal from Zden{\v e}k, and our research paths intersected frequently.
Our names appear together on only 12 papers, though, notably the first of his series ``Large-Scale Active Coronal Phenomena in \textit{Yohkoh}~SXT Images,'' with which he brought his \textit{Skylab} experience to bear on data that I was closely familiar with.
See Section~\ref{sec:arches} below for my views of how this developed.

\section{Optical Spectroscopy of Flare Emissions}\label{sec:wlf}

\subsection{Spectroscopy then}

Zden{\v e}k's research career began with optical spectroscopy of solar flares -- the tools available at the time permitted high spectral resolution and therefore supported analyses of radiation physics in the flaring medium, but these tools were cumbersome.
A good observation typically produced a well-defined spectrum with not-so-good seeing, limited metadata, and perhaps another one or two at different phases (and locations of the long slit) on the flaring region.
Typically the observer would place the slit at the brightest point of the flare and be thankful if good signal-to-noise properties developed.
Zden{\v e}k worked extensively in this area and led the development of flare spectroscopy at Ond{\v r}ejov observatory in the then Czechoslovakia.
This work taught us a great deal about the behavior of the lower solar atmosphere during a flare \cite[\textit{e.g.}][]{2007ASPC..368..387B}, and also about the ``loop prominence systems'' (see below), which interestingly contained low-temperature gas embedded magnetically in the hot corona.
It also led to review papers \citep{1966SSRv....5..388S,1972ARA&A..10....1S} and to his monograph ``Solar Flares'' \citep{1976sofl.book.....S}.
This book remains current, perhaps partly because the attention of solar-flare researchers turned away from optical ground-based observations and the complexities of the atmospheric structures they reveal, as the new spectacular observations from space developed -- for example, with the \textit{Skylab} observatory, which boosted Zden{\v e}k into a new career phase as described below.

\begin{figure}[htbp]
\centering
   \includegraphics[width=0.49\textwidth]{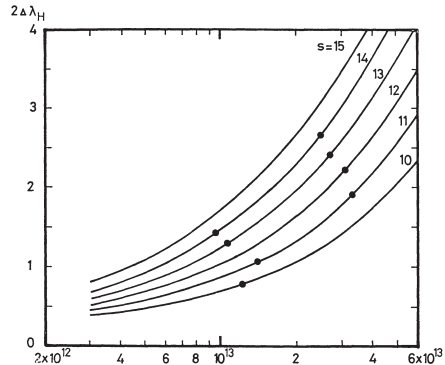}
     \caption{Balmer-series observations of two major flares (SOL1956-12-29 and SOL1960-08-07), showing resuts as two sets of black dots.
     These correspond to mean densities $n_e = 1.2 \times 10^{13}$ and $2.9 \times 10^{13}$~cm$^{-3}$, respectively.
       }  
\label{fig:halpha_density}
\end{figure}

High-resolution optical spectroscopy thus led to many insights in flare physics.
Chief among Zden{\v e}k's particular contributions was the work on Stark broadening in the profiles of the Balmer-series lines.
These showed that high densities must prevail \citep{1959PASJ...11..185S,1972ARA&A..10....1S}; estimates from that era put $n_e > 10^{12}$~cm$^{-3}$ for flares observed against the disk.
We readily understand this as regards the flare ribbons, embedded in the dense chromosphere, and the direct implication is that the chromosphere has been substantially ionized by the flare process, when compared with models of the quiet chromosphere.
Figure~\ref{fig:halpha_density} shows representative observations of this effect.

For limb flares, we are looking at the mis-named \citep{2007SoPh..246..393S} ``loop prominence systems,'' to be discussed in more detail in Section \ref{sec:lps} below along with their mis-namings at other wavelengths.
These structures may also have high densities, such that with high spatial resolution one can see emission in the core of H$\alpha$ even when the coronal structures are projected against the solar disk.
\cite{1983SoPh...86..185H} found a coronal electron density $n_e \approx 10^{12.8}$~cm$^{-3}$ for the limb flare SOL1982-06-12.

\subsection{Spectroscopy now}

Where has the optical spectroscopy of solar flares gone since Zden{\v e}k's time?
Ground-based observations have been hit-or-miss; we could point to the multi-slit spectra (with film readout) of \citep[\textit{\textit{e.g.}},][]{1974SoPh...37..343M} or the initial CCD-based observations of \cite{1987SoPh..114..115W}.
Such observations struggled with the basic requirement for imaging spectroscopy at simultaneous (and high) spatial, temporal, and spectral resolution. 
We note also that observations of stellar flares, though without the advantage of spatial resolution, offer different views of probably the same physics. 
``Sun-as-a-star'' observations of high-resolution optical spectra also can be achieved with instruments not originally designed for solar work \citep[\textit{e.g.}][]{1997ApJS..112..221J}; such observations make use of cross-dispersed echelle spectrographs/sp with broad spectral coverage.

It would be fair to say that observations of the lower solar atmosphere during flares, enabled by the advanced kinds of imaging spectroscopy using current technology, are only just returning to the mainstream of research into solar flares.
In the meanwhile the advance of technology has made stellar spectroscopy immeasureably better, and although this may seem embarrassing (why not study the nearest star most diligently?) the stellar observations provide a wonderful complement.
The period of slower development in the meanwhile has an obvious explanation: access to space made it possible to do absolutely new things, and at the same time it was clear to all that the lower solar atmosphere presents huge problems for model development.

\section{``Proton flares''}\label{sec:proton}

\subsection{Proton flares then}

The phenomenon of the ``proton flare'' refers to the acceleration of high-energy particles (protons above 10~MeV) at or near the Sun, which then appear near the Earth as ``solar cosmic rays.''
The presence of such particles definitively establishes the fundamental non-thermality of flare processes, since we typically observe energetic protons at 10 MeV, some 10$^5$~kT in terms of a coronal source.
We now term these particles SEPs (Solar Energetic Particles) and recognize a huge variety of properties in terms of elemental abundances, ionic states, and even isotopic composition.
In the 1950s it had already been well-established that there were often strong connections between sunspots (or flares) and geomagnetic storms \cite[see, for example][]{chapman-bartels}.

In the 1950s the problem therefore was to establish observational signatures of the extreme particle acceleration that could happen in association with flares.
This presumably was what led Zden{\v e}k to creating an atlas of proton flares from 1938 to 1955, detected as the ``Polar Cap Absorption'' (PCA) signature in ionospheric soundings.
From 1942, neutron monitors entered the scene, and of course the remarkable event SOL1956-02-23 \citep{1956PhRv..104..768M}.
This just escaped Zden{\v e}k's catalog, but in several subsequent papers he explored the statistical aspects of the proton morphology in the context of the flare process.
Eventually he helped to organize a ``Proton Flare Project'' of dedicated observations, as reported by \cite{1969SoPh...10....3S}.
Although many technical conclusions appeared, there was no obvious success at achieving a basic physical understanding of how flares could accelerate all of these particles.

\begin{figure}[htbp]
\centering
   \includegraphics[width=0.58\textwidth]{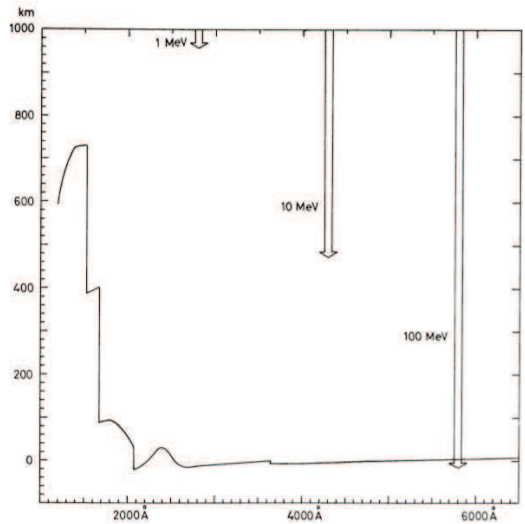}
     \caption{An early view of the thick-target model for solar flares, showing height in km \textit{vs}. optical wavelength in the photosphere and low chromosphere.
     The solid line shows the location of unit optical depth as a function of wavelength, and the arrows show the penetration of vertically-incident protons accelerated in the corona. 
     Protons require of order 100~MeV to reach the visible photosphere.
       }  
\label{fig:zdenek_ttm_cartoon}
\end{figure}

The relationship between the particles detected in interplanetary space and those required for flare development also remained murky, although at this early time it was by no means clear that particle acceleration played the central role in flare physics that we now recognize.
Nevertheless Zden{\v e}k created the cartoon shown in Figure~\ref{fig:zdenek_ttm_cartoon} \citep{1970SoPh...13..471S}, showing that protons could in fact heat the lower solar atmosphere -- this was not a new result, but the analysis was excellent.
He also went on to provide estimates for the numbers of energetic neutrons directly arriving at Earth from a flare \citep{1971SoPh...19..202S}, a phenomenon predicted by \cite{1951ZNatA...6...47B}, elaborated by \cite{1965JGR....70.4077L}, and subsequently discovered by \cite{1982ApJ...263L..95C}.

The impulsive phase of the flare, however, has a close identification with the bremsstrahlung emission of 10-100 keV electrons, not protons as suggested by the ``proton flare'' association and Figure~\ref{fig:zdenek_ttm_cartoon}.
This change resulted in the ``thick-target model'' of an \textit{electron beam} originating in the corona, rather than a \textit{proton} beam \citep[\textit{\textit{e.g.}}][]{1971ApJ...164..151K,1971SoPh...18..489B,1972SoPh...24..414H}.
It seems unlikely now, for various reasons, that Zden{\v e}k's proton beam can play a role in the impulsive phase, or if it does then the proton acceleration would likely be in the lower atmosphere directly because of the timescales involved.

\subsection{Proton flares now}

After the SOL1956-02-23, still ranking as the most powerful SEP event on the records, space research developed rapidly and flare-associated particle events have now become a major component of ``space weather.''
Nevertheless confusion still exists about the role of flares in SEP production, even though comprehensive evidence exists establishing that high-energy particles do indeed exist in flare plasmas.
The acceleration of SEPs has a clear relationship with the interplanetary propagation of large-scale shock waves driven by CME expansion \cite[\textit{\textit{e.g.}}][]{2013SSRv..175...53R}, and so these particular particles do not have any obvious association with those responsible for either the bremsstrahlung hard X-rays or the $\gamma$-ray emission by solar flares. 
For both of these components there is strong evidence for the non-thermal action of particle acceleration to take place on closed magnetic-field structures in the core of an active region, rather than in large-scale coronal structures involved with the CME \citep[\textit{e.g.}][]{2011SSRv..159...19F}.

Recently modern data have jeopardized the thick-target paradigm (powerful electron beams); \cite{2012ApJ...753L..26M} successfully made absolute height determinations for the hard X-ray sources in the white-light flare SOL2011-02-24 and found them to be at the photosphere, out of the range of any plausible electron beam.
At present we have no good working model for the process: neither a {\u S}vestka-style proton beam nor an electron beam seem to be appropriate.

\section{Loop Prominence Systems}\label{sec:lps}

\subsection{Loop prominences then}

One of Zden{\v e}k's earliest papers dealt with observations of solar flares in H$\alpha$, as a part of his work at Ond{\v r}ejov in the group of Prof. F. Link \citep{1949BAICz...1...73S}, and with a publication date 58 years later, he concluded his independent research activities with a comment on the nomenclature and physics of the coronal mani\-festations of solar flares \cite{2007SoPh..246..393S}.
Both of these papers dealt with the coronal effects of solar flares.

\begin{figure}[htbp]
\centering
   \includegraphics[width=0.58\textwidth]{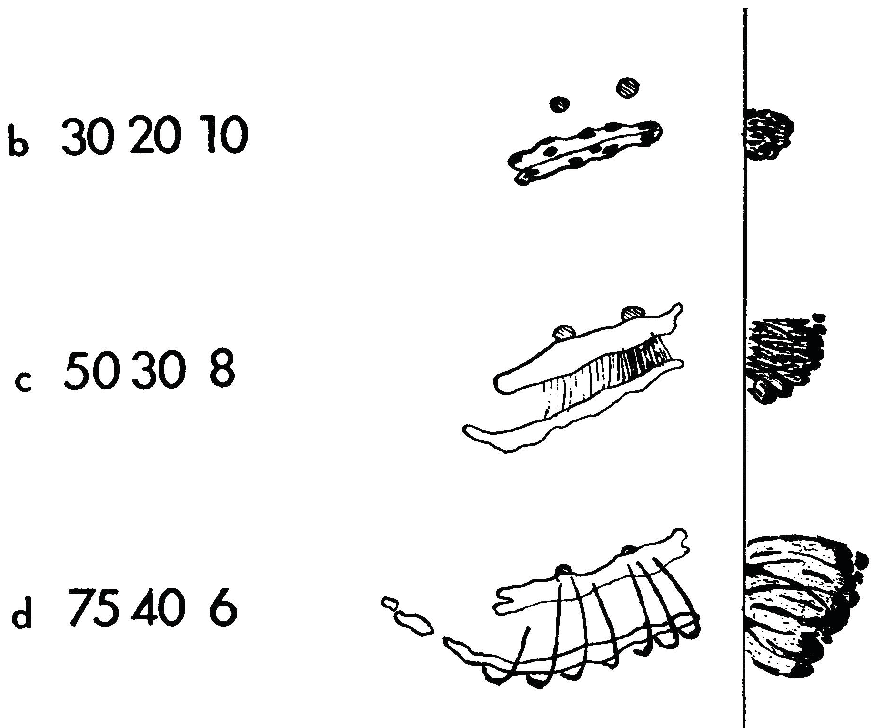}
   \includegraphics[width=0.40\textwidth]{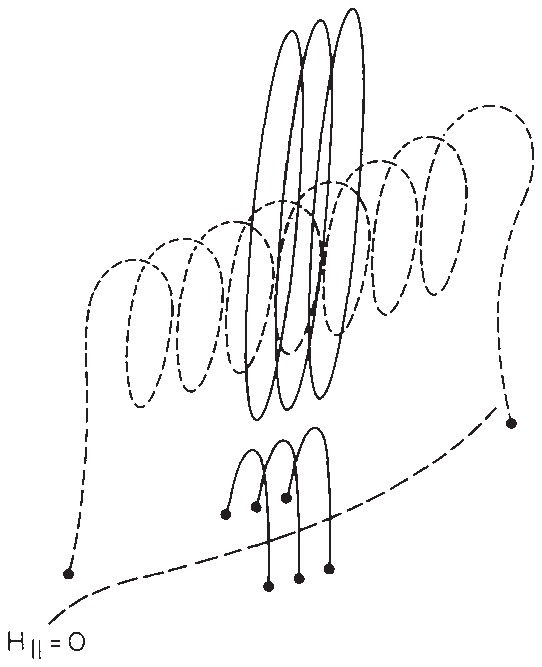}
     \caption{Left, adapted from Bruzek (1964), sketching 
     H$\alpha$ loop systems for three flares as they would be seen on the disk, and off the limb.
     The numbers in the legend show minutes after onset, height in~Mm, and upward velocity in km/s.
     Right, from \cite{1983SSRv...35..259S}; this illustrates how 2D reconnection would simply create coronal toroids, whereas 3D reconnection could produce a flux rope.
       }  
\label{fig:bruzek}
\end{figure}

The loop-prominence phenomenon played a major role in laying the framework of our current understanding of flare processes in the solar corona. 
Before Zden{\v e}k's time one had the clear knowledge that loop prominence systems had a strong association with the occurrence of major solar flares, and \cite{1964ApJ...140..746B} provided the now-iconic view (\textit{e.g.}, Figure~\ref{fig:bruzek}, left panel) of the growth of large systems of coronal loops, now termed ``arcades,'' which H$\alpha$ observations revealed  most clearly when they rose high enough to appear above the limb.
The Bruzek paper also firmly established the existence of a relationship between the proton flares and the formation of a loop-prominence system.
We can also nicely trace the ideas arising from these flaring loops, and their ribbon-like footpoints, to the popular flare model of large-scale magnetic reconnection now termed the ``CSHKP'' model (Carmichael, Sturrock, Hirayama, Kopp-Pneuman); Zden{\v e}k was captivated by the description \cite{1976SoPh...50...85K} gave, but he recognized (Figure~\ref{fig:bruzek}, right panel) the need for three-dimensionality.

The coronal development of an eruptive flare has many signatures, ranging actually across the entire electromagnetic spectrum that is accessible to us.
Accordingly, many misleading neologisms have popped up.
In the coronal optical emission lines (\textit{e.g.}, the ``green line'' of Fe~{\sc xiv} at 5303~\AA, but especially the ``yellow line'' of Ca~{\sc xv} at 3594~\AA) spawned the terminology of the ``sporadic coronal condensation,'' and the microwave observations created the terminology ``post-burst increase.''
There has recently been some use of the term ``post-eruption arcade'' (PEA), even though an arcade can appear without any eruption
\citep[\textit{e.g.}][]{1998A&A...332..353G}.
Zden{\v e}k felt that ``post-flare loops,'' and presumably all of these others, might more efficiently just be called ``flares'' \citep{2007SoPh..246..393S}.

\subsection{Loop prominences now}

The MHD modeling of ``eruptive flares'' -- including those with CMEs, has become a major activity with the recognition of the importance of ``space weather'' and the need to anticipate or predict the occurrence of extreme events capable of major damage to human society \citep{2013SpWea..11..585B}.
The sketch of data and the ensuing cartoon description (Figure~\ref{fig:bruzek}) have evolved into a major activity involving many research workers and participating agencies, including commercial users of space-weather data and models.

As the solar aftermath of an eruption (though not uniquely so, as noted above) a set of flare loops usually provides an excellent guide to the heliographic region underlying a CME event.
Usually powerful events have associations with active regions, but sometimes not -- filament eruptions in quiet regions (\textit{disparitions brusques}) may have effects strongly resembling those of active-region flares  \citep[\textit{e.g.}][]{1986stp..conf..198H}.
Such eruptions may involve coronal footprints much larger than those of active-region eruptive flares \citep[\textit{e.g.}][]{1995JGR...100.3473H}, even if the ensuing CME may occupy a similar fraction of the heliosphere.
These events have a shadowy research history since they do not show up readily in GOES soft X-rays, never achieving high temperatures, and overlap in parameter space with the few cases of ``stealth CMEs'' that have virtually no detectable low-coronal counterpart \citep[\textit{e.g.}][]{2009ApJ...701..283R}.

The loop prominence systems, in modern data, follow the pattern that led to the development of the standard reconnection models.
A large two-ribbon flare consists of an orchestrated array of individual loops; each (according to the accepted picture) forms via large-scale reconnection.
The energy newly deposited in an elemental newly formed loop renders it under-dense and over-pressure in terms of hydrostatic equilibrium,  so that plasma flows from the chromosphere below must try to restore this equilibrium as the loop cools.
Eventually the temperature drops to the point of Field's thermal instability \citep{1971SoPh...19...86G}, condensation ensues, and ``coronal rain'' results when H$\alpha$ and other chromospheric emissions suddenly appear out of nowhere and slide down along the field structure.
Recently \cite{2014ApJ...780L..28M} discovered that the \textit{Helioseismic and Magnetic Imager} on SDO can readily detect major arcade systems in the continuum near the magnetic line 6173~\AA.
These observations have high spectroscopic resolution (though only six points over a narrow band) and full polarization capability, exploited by \cite{2014ApJ...786L..19S} to show that an appreciable component of the continuum emission at this wavelength is in fact due to electron scattering.
The observations also reveal the presence of post-flare coronal rain as expected.
This serendipitous capability, plus the wealth of coronal information from AIA, should allow future research to understand the energetics
of these arcade structures more precisely.
Section~\ref{sec:arches} contains some related information.

\section{The Flare Build-up Study}\label{sec:fbs}

\subsection{Flare build-up then}

The Flare Build-up Study (FBS, organized by Zden{\v e}k and Paul Simon) began with a workshop in 1975 and continued until 1986 \citep{1987SoPh..114..389G}.
Its premise was the need for a proof of the concept that flare energy could be stored magnetically in the solar atmosphere, and to find out how it could be released.
This large-scale organization involved on the order of a hundred scientists and dozens of participating observatories, with several workshops during the course of the activity; important programs involved included the \textit{Solar Maximum Mission} spacecraft among others, the \textit{Very Large Array} and other radio facilities, the Kitt Peak Observatory among several ground-based observatories, and
the initial coordinated observations focused on NOAA active region 5022 during western-hemisphere daylight hours in June, 1982.

Indeed, much data resulted and provided material avidly discussed during the workshops that ensued.
\cite{1989SoPh..121..135G} provided a general and still-current description of preflare activity based upon these results, for example.
The topics addressed sound familiar now, three decades later: flux emergence, shear, magnetic complexity, flare homology, magnetic reconnection, large-scale photospheric magnetic patterns, not to mention flux cancellation at the small scales.
At this time the initial difficult observations of the vector magnetic field were just appearing (Hagyard et al., 1982; see also Severny, 1964) and there were no systematic soft X-ray or EUV coronal images with which to define the coronal geometry in adequate detail.
\nocite{1964ARA&A...2..363S}
\nocite{1982SoPh...80...33H}

\subsection{Flare build-up now}

In retrospect the Flare Build-up Study did not provide decisive quantitative answers to the key questions: (i) Does the coronal field uniquely store the flare energy prior to its release,  (ii) Can we establish a ``relaxation oscillator'' (loading/unloading) behavior that directly links energy input and release, and (iii) What are the basic instabilities of the structure that create the flare?
The latter two of these questions (not exactly as asked by the FBS) still remain open to this day, but we can recognize clear signs of progress. 
As Zden{\v e}k and the organizers of these early campaigns realized, the data of that day were incomplete -- but were they inadequate?
Perhaps not, because flares inevitably involve large scales even if the unobservable microphysics does not.
So in my view the main problem was, and remains, theoretical understanding.
Things have improved enormously, but still with no convergence on the answers to questions of this sort.
In the following I comment on each of these items in turn.

(i) We have seen vast improvement in our ability to characterize the coronal magnetic field based on steadily-improving photospheric observations of the vector field, including much sharper ways of assessing the free energy available for release in a flare.
Not only do these extrapolations reveal the basic requirement, namely the storage of sufficient energy for a major flare, they also in some sense localize it \cite[\textit{e.g.}]{2009ApJ...696.1780D,2012LRSP....9....5W}.
In many cases now, these methods actually show a time-series decrease in the total coronal magnetic energy of an active region at the time of a major flare \citep[\textit{e.g.},][]{2012ApJ...748...77S}.

(ii) In a relaxation oscillator, a steady input of energy into a system that can become unstable results in characteristic forms of time structure.
If the trigger mechanism of the instability is itself well-determined, as in an electronic circuit, then the time structure is periodic.
If not, then there is a proportionality between the energy release of the instability and the time interval before, or after \citep[\textit{e.g.}][]{1998ASSL..229..237H}, the trigger action.
If the energy input is not steady, the pattern becomes correspondingly confused. 
Many natural systems show this kind of behavior \citep[\textit{e.g.}][]{1976ApJ...207L..95L,2006ApJ...652.1531M} -- perhaps even the human heartbeat!

The Flare Build-up Study led to searches for this kind of effect \citep{1982AdSpR...2...11G}, making use of sequences of homologous flares where one should have the best opportunity to detect a pattern, on the plausible assumptions that flux emergence has some persistence and that the homology of the resulting flares indicates a consistent trigger threshold.
This study did not provide convincing evidence of the expected pattern, but the several assumptions required offer possible explanations.
Most obviously the lack of a correlation could reflect the irregularity of the input of free energy from the photosphere into the corona, closely related to possibly complicated pattern of emerging flux; how persistent is it?
In addition we still do not understand the physical mechanism of the flare instability, nor understand how it may be triggered.
Nevertheless the detection of a significant correlation would confirm the simplest picture of the process.

(iii) The identification of which kind of instability can produce a flare remains contentious. 
We have seen one major development since 1986, in that the notion of ``shear'' has turned into a substantial effort to understand the description of \textit{helicity} in the magnetized coronal plasma and its role in defining the accessible stored energy, and in linking flares with global field restructurings such as flare, filament eruptions, and CMEs \citep[\textit{e.g.}][]{1994GeoRL..21..241R,2001JGR...10625141L}.

\section{Giant Arches}\label{sec:arches}

\subsection{Giant arches then}
The loop prominence systems identified with CMEs had suggested the CSHKP model of magnetic reconnection, and the data from \textit{Skylab} (1973-74) and the \textit{Solar Maximum Mission} (SMM; 1980-89) really solidified these ideas.
Zden{\v e}k participated in data analysis from both of these missions, and his ideas about ``giant arches'' began with analysis of a flare well observed by SMM and radio observatories \citep{1982SoPh...75..305S}: SOL1980-05-21 (X1), the same event from which \cite{1981ApJ...246L.155H} observed the relationship between hard X-ray footpoint sources and flare ribbons.
Both of these observational results derived from hard X-ray observations by the HXIS instrument on board SMM, a remarkable hard X-ray telescope \citep{1980SoPh...65...39V} based on innovative technology never again utilized.
The hard X-ray imaging results from this instrument, both in the lower solar atmosphere and high up in the corona, anticipated many of the results now associated with the \textit{Yohkoh} and RHESSI hard X-ray imaging.

\begin{figure}[htbp]
\centering
   \includegraphics[width=0.43\textwidth]{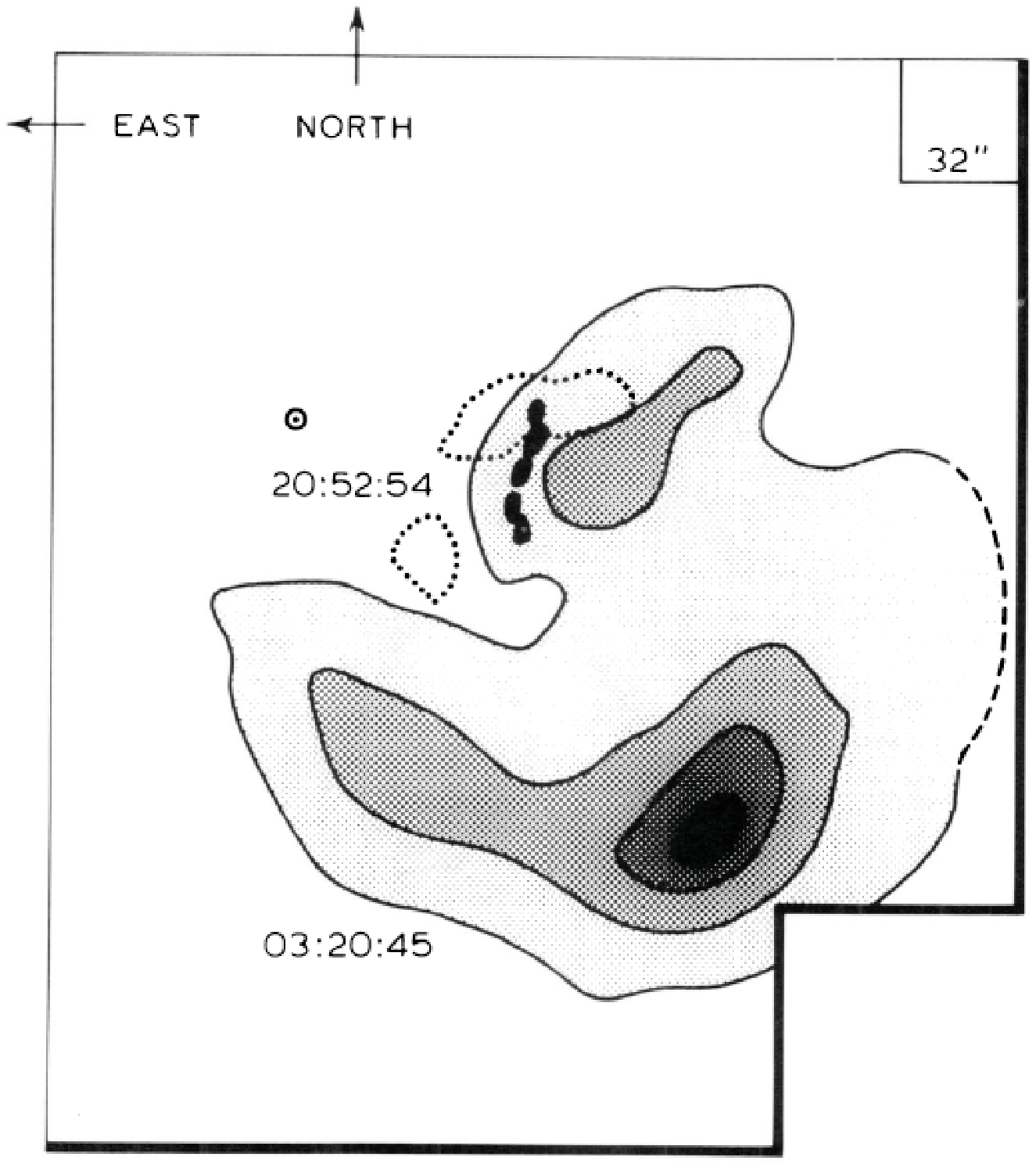}
   \includegraphics[width=0.55\textwidth]{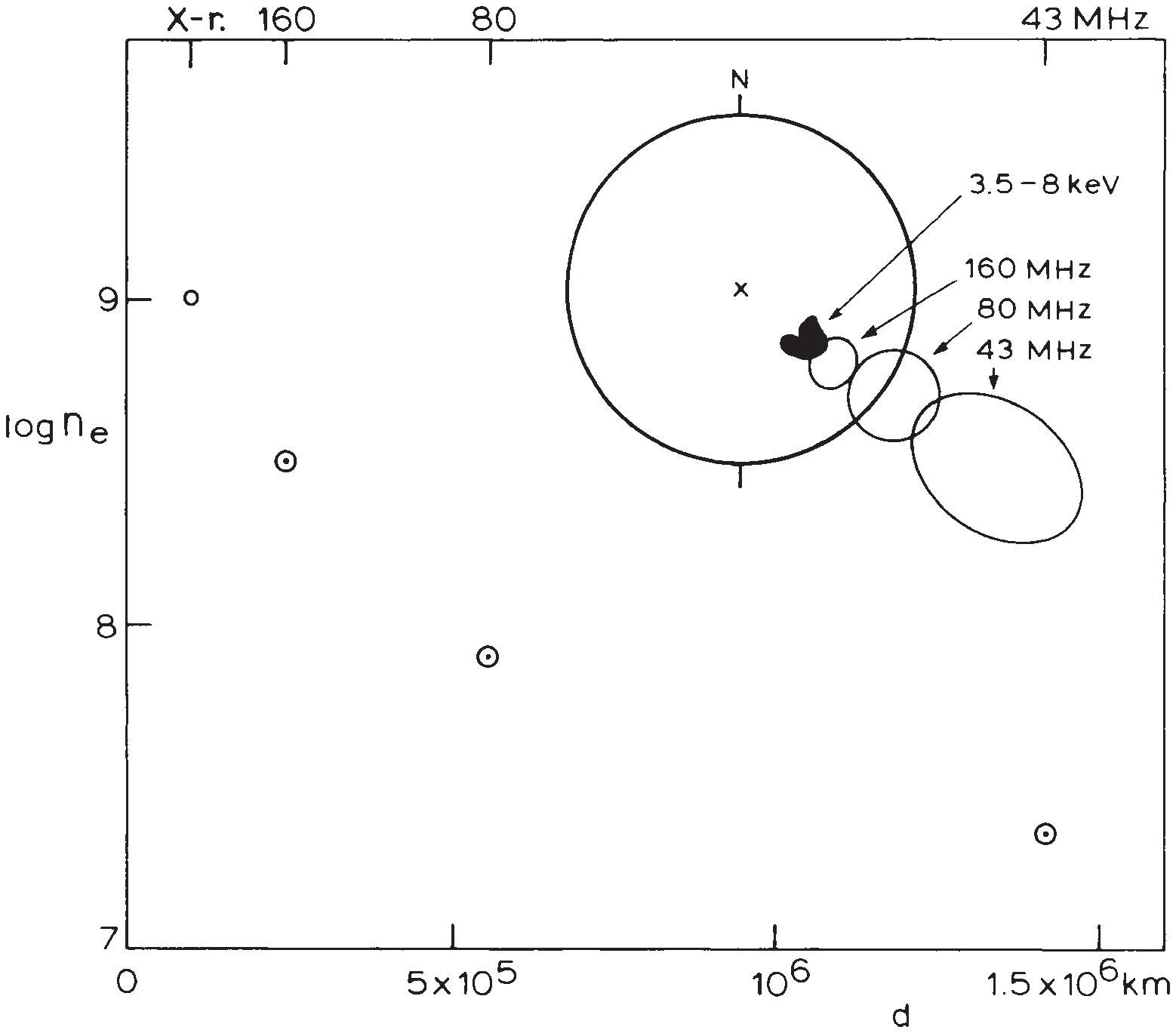}
     \caption{Left: the long-lasting coronal arch structure observed by HXIS as a late development of SOL1980-05-21 (energy 3.5-5.5~keV).
     The odd shape of the field of view is a part of the design of this instrument; the box at upper right shows its 32$''$ pixel size for this observation.
     Right, observations  of the same event at several frequencies from the Culgoora radio telescope.
     The points (axis to left) represent density estimates for the sources seen at the three Culgoora wavelengths and in HXIS soft X-rays.
     Figure adapted from \cite{1982SoPh...75..305S}.
       }  
\label{fig:arch}
\end{figure} 

The description ``arch'' obviously leaps to mind from Figure~\ref{fig:arch}, from \cite{1982SoPh...75..305S}.
The paper argued convincingly that this arch could be distinguished from the post-flare arcade, and had a separate dynamical development in the outermost part of the active region.
The structure seen in X-rays and in meter-wavelength images from Culgoora (right panel) remained static, rather than growing with time as expected for reconnection products.
This structure could thus not readily be understood in terms of the \cite{1976SoPh...50...85K} picture from this and from points of view described in this paper.
This paper attracted some interest, but it would be fair to say that much of the community remained unconvinced that anything beyond the Kopp~and Pneuman model would be needed.

\textit{Yohkoh} arrived,  carrying the first CCD-based soft X-ray telescope with good resolution in space and good sampling in time.
We immediately could get a better view of what SMM had glimpsed, and Zden{\v e}k eagerly wrote a series of papers on the flare arcades and giant arches \citep{1995SoPh..161..331S,1996mpsa.conf..609S,1996SoPh..168..331F,1997SoPh..176..355S}, specifically showing that the new and better soft X-ray imaging confirmed the existence of the somewhat shadowy ``giant arches.''
In the final paper of this series we see the suggestion that large-scale reconnection was happening on two scales.
The event analyzed in the paper, SOL1992-10-28T10:21 (C3.1), illustrated this well.
The discussion suggested that the reconnection leading to the re-formation of the coronal streamer could be identified with the giant arches, and that the process leading to the flare loops proceeded independently, even though both sets of phenomena appeared in large-scale two-ribbon flare events.
\cite{1990SoPh..127..267K} provide a full discussion of this distinction and what it implied within the framework of the HXIS-era data \citep[but see also][]{2015ApJ...801L...6W}.

\begin{figure}[htbp]
\centering
   \includegraphics[width=0.6\textwidth]{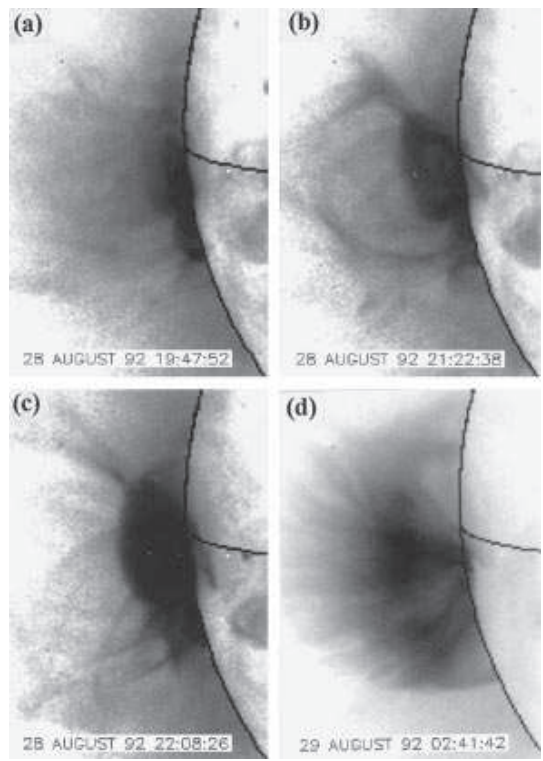}
     \caption{SOL1992-08-28 (C1.5), a beautiful flare event at the limb that displays the hot fan-like structure studied in \cite{1998SoPh..182..179S}. 
     The four panels show soft X-ray emission at various stages of the event, as described in the text.
     The hot, spiky fan structure in the late image (02:41~UT, some five hours after the initiation of the event) is remarkable.
       }  
\label{fig:SOL1992-08-28}
\end{figure}

A final paper in Zden{\v e}k's series of papers entitled ``Large-scale active coronal phenomena in \textit{Yohkoh} SXT images'' series dealt with the hot fan-like structure of the late phase of the limb event SOL1992-08-28 (C1.5), as illustrated in Figure~\ref{fig:SOL1992-08-28} \citep{1998SoPh..182..179S}.

Events of this kind contain the dynamical phenomena we term ``supra-arcade downflows'' \cite{1999ApJ...519L..93M}.
These could only have been discovered via \textit{Yohkoh}/SXT observations by a revised pattern of exposures not implemented until 1998,
but their discovery dramatically confirmed the presence of dynamics closely resembling that expected of a global reconnection model such as that of Kopp~and~Pneuman.

The soft X-ray images of the Sun also revealed large loop structures interconnecting active regions, including the interesting trans-equatorial loops. 
Because of the large scale of these structures and their association with flare or CME activity, they had interest in the context of the giant arches \citep{1995SoPh..161..331S,1999SoPh..187...33F}.

\subsection{Giant arches now}
Zden{\v e}k's active period of research came to an end just a decade or so too soon; the voluminous high-resolution data from the \textit{Solar Dynamics Observatory} has clarified much of what he inferred from much-inferior data.
We note that these data have wavelength and temperature sensitivities that are different from the soft X-ray observations of SMM/HXIS, \textit{Yohkoh}/SXT, or \textit{Hinode}/XRT, but they have wonderful coverage, angular resolution and continuity.
We note two important developments that he just missed: the ``supra-thermal arcade downflows''  \citep[SADs;][]{1999ApJ...519L..93M} and the 
``EUV late phase'' with its associated morphology \citep{2011ApJ...739...59W}.
Both of these concern our interpretation of the giant arches.

\subsubsection{SADs}
Exactly the spiky fan structures  seen in soft X-rays (Figure~\ref{fig:SOL1992-08-28}) turned out to support a characteristic pattern of downflows in the form of dark blobs seen in soft X-ray images \citep{1999ApJ...519L..93M}.
These structures occupy the region where a reconnection outflow jet would appear in the standard model, but have sub-Alfv{\' e}nic speeds and a gradual deceleration, features not predicted either in the cartoon or in related MHD simulations of large-scale reconnection processes in solar flares \cite[\textit{e.g.}][]{2011LRSP....8....6S}; such models also predict that the reconnection outflow would terminate in a fast-mode MHD shock, which is not observed.
Thus the SADs, and the spiky arcade tops described by \cite{1998SoPh..182..179S} simultaneously offer striking evidence of large-scale magnetic reconnection, upon very complicated separatrix structures, but also have not yet been convincingly modeled.

The SADs can originate in the corona well above where the original soft X-ray observations lost sight of them;
\cite{2004ApJ...616.1224S} could track the flows out into the domain of LASCO.
Their presence in the high corona might make them a candidate for the modern identification of the phenomena that Zden{\v e}k called ``giant arches'' in some sense, but current thinking associates them with the main flare eruption and the restructuring following it: the flare loops.
Recent work on SADs suggests that the whole concept of coronal magnetic reconnection at low plasma beta may need revision \citep[\textit{e.g.}][]{2001ApJ...557..326S,2013ApJ...766...39M}.
This radical change in our thinking about the microphysics may fit comfortably enough within our well-developed observational understanding of the magnetic restructuring involved in a major eruptive flare; in other words, Zden{\v e}k may not have been too surprised by this interesting possible revision.

\subsubsection{EUV late phase}
The ``EUV late phase'' phenomenon \citep{2011ApJ...739...59W} appeared first in the Sun-as-a-star spectroscopy from the EVE instrument on SDO.
These data, though simple EUV spectra, are sensitive and precise.
Furthermore the AIA instrument on SDO allows us to identify the spectral variations with image features \citep{2014SoPh..289.3391W}, which include eruptive signatures as detected directly in image motions as well as via coronal dimmings \citep[\textit{e.g.}][]{2001JGR...10625199H}.
These data showed clearly that in addition to the  flare disturbance as seen by GOES at high temperatures (above Fe~{\sc xx}, say), sometimes other disturbances followed as a part of the late development of the overall event.
These later disturbances typically appear most prominently at Fe~{\sc xvi} -- elevated temperatures, but not as high as in the impulsive or gradual phases previously known.
They involve large-scale loop structures not directly associated with the flare gradual phase, and \cite{2013ApJ...768..150L} note that a CME may or may not accompany an EUV late-phase event.
As a speculation, we suggest the identification of the EUV late-phase events with Svestka's giant arches.

\section{Conclusion}

In this paper I have reviewed some of Zden{\v e}k {\u S}vestka's most interesting research activities, in some of which I was fortunate enough to help out.
It is quite remarkable how much progress has ensued in the recent few years, and I have tried to point out where Zden{\v e}k's discoveries and insights may fit into the new context.
Of course, during his day the data were hugely less competent, and we could only get glimpses of phenomena now laid out for us in striking detail  -- if sometimes still perplexing. 
It is most unfortunate that Zden{\v e}k did not see some of these developments, especially those which bore out some of his favorite ideas.

The optical spectroscopy of solar flares, especially white-light flares, remains as a lasting legacy of Ond{\v r}ejov observatory, and Zden{\v e}k in particular. In 1859 this phenomenon opened the door for many kinds of research, including the current major developments in ``space weather.''
We should remember that this original flare predated both Heaviside and R{\" o}ntgen, whose discoveries later in the 19th century led eventually to an understanding of direct solar flare effects.
Zden{\v e}k's research work began almost halfway in between then and now.


%

%

%

%
 \begin{acks}
This work was supported by NASA under grant NNX11AP05G for studies of white-light flares and by contract No. NAS 5-98033 for RHESSI at UC Berkeley.
I would like to thank the University of Glasgow for hospitality during the preparation of this paper, and several friends for comments on the first draft of the paper: Franta F{\' a}rn{\' i}k, Vic Gaizauskas, and Giannina Poletto.
 \end{acks}

%
%
\bibliographystyle{spr-mp-sola}
\bibliography{svestka}  
%
%
%
%

\end{article} 
\end{document}